\documentclass[a4paper,11pt]{article}
\usepackage{pos}
\usepackage{subcaption}
\usepackage{hyperref}
\usepackage{placeins}                    %
\usepackage[sort&compress]{natbib}

\title{PeV neutrons as origin of separated SS433 TeV signals}

\author*[\orcid{0000-0003-3146-3932}, a,b]{D. Fargion}
\author[\orcid{0000-0001-7503-2064}, c]{P.G. {De Sanctis Lucentini}}
\author[\orcid{0000-0002-4603-8405}, d]{S. Turriziani}
\author[\orcid{0000-0002-1653-6964}, e]{M.Y. Khlopov}
\author[f,g]{D. Sopin}
\affiliation[a]{Rome University “La Sapienza” and MIFP, Rome, Italy.}
\affiliation[b]{Osservatorio Astronomico di Capodimonte, INAF, Naples, Italy}
\affiliation[c]{Gubkin University, Moscow, Russia.}
\affiliation[d]{Centro de Astronomía (CITEVA), Universidad de Antofagasta, Av. Angamos 601, Antofagasta, Chile}
\affiliation[e]{Virtual Institute of Astroparticle physics, 75018, Paris, France}
\affiliation[f]{Institute of Physics, Southern Federal University, Stachki 194 Rostov on Don 344090, Russia}
\affiliation[g]{National Research Nuclear University MEPhI, 115409 Moscow, Russia}

\emailAdd{daniele.fargion@fondazione.uniroma1.it}

\abstract{
The SS433,  a well-known binary system with an internal black hole, have shown since half a century, an inner (a few year light distances) twin precessing jets spirals. These beams are  made by tidal forces  while stripping mass from large stellar companion feeding an inner BH accretion disk and an orthogonal accelerating twin jet. From it, the radio, X gamma jet emission. A couple of years ago H.E.S.S telescope as well as  HAWC and LHAASO array detectors, discovered also the surprising signature of an unexpected   far twin separated gamma beam at tens TeV energy. At  a hundred light years distances from its central source. 
We suggest that it is the legacy of  a past rare eruption, a century ago,  of tens  PeV  ($10^{16} $ eV) relativistic twin  neutron  beams. Their  beta decay in flight at far distances, into proton, neutrino and in particular into tens TeV electrons, could feed the observed TeV gamma traces. They are originated by the same secondary tens TeVs electrons  emitting  hard gamma,  by Inverse Compton Scattering  onto interstellar infrared photons.
}

\FullConference{Multifrequency Behaviour of High Energy Cosmic Sources - XV (MULTIF2025)\\
 9-14 June, 2025\\
Mondello, Palermo, Italy\\}

\graphicspath{{Figs/}{./}}
\def\orcid#1{\kern .08em\href{https://orcid.org/#1}{\includegraphics[keepaspectratio,width=0.6em]{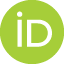}}}
\begin{document}
\maketitle

\section{Introduction}
SS433 is one of the first micro-quasars to be discovered and understood as a binary system. A ten solar mass Black Hole, BH, and a comparable companion mass, a large heavy star in tidal nearby orbit. In all micro-quasars  the wide mass of the star is being  stripped by the compact  BH ( or Neutron Star  NS). This process typically feeds an accretion disk around the NS or BH.  The same disk by its net separated charged rings, while spinning, produces huge magnetic fields  $ B_d$. Their time variability ${dB(t)/dt}$ induce electric fields capable of accelerating charges in the relativistic regime,  These charge flow are collimated by the disk magnetic field  into twin jets escaping orthogonal to the disk plane. These jets are tidally deflected by the companion star mass,  into a twin, precessing jet. This is the inner nature of the micro-quasar as an SS433 system.
The spiral tail of the SS433 jets is showing the precession of an ultra-relativistic outflow, spraying nucleons and also electrons at relativistic speeds,  see Fig.\ref{fig:1}. The twin  jet has been observed in radio, X-ray, and gamma-ray spectra since 1978-1979. Its long spiral tails are diffuse and diluted over a distance of a fraction of a light-year (ly).  The source SS433 lies inside  the old ($18.000-20.000$ years ago) supernova remnant nebula W50, whose small asymmetry reveals the jet's past and present role. Very recently, H.E.S.S.~\cite{hess2024acceleration} and HAWC~\cite{alfaro2024spectral} surprisingly discovered the appearance of a twin hard gamma-ray  beam, very distant and well disconnected from  the SS433's inner jet. This separated twin beam-jet  at  a distance of about 75 or even up to 150 light-years from the same inner SS433 source is very surprising.
A recent model is trying to explain the observations, based on an accelerating shock wave (at a very far distance) that will re-accelerate up to PeV energies the beam of nuclei and their lepton TeV secondary tracks. The observed collimation of the PeV-TeV beam jet is very difficult to accept in  this model.
We prefer an alternative model based on known high-energy nuclear physics: the burst of a  tens PeV ($ > 10^{16}$ eV) neutron  beam and its  ultra-relativistic beta decay in flight.
That neutron beam  may occur when a proton jet of tens of PeV is converted by photon-pion reactions,  to a neutron jet of comparable  tens of PeV energy.  %
This may have occurred in the past due to a hot nova-like  flare or burst \cite{bernabei2025proceedings,fargion2024ss433,Fargion:20259k}.

This neutron beam, ejected within energy spectra around tens of PeV ($10^{15}$ eV) range, may offer the ideal courier for the TeV far twin gamma traces. The ultra relativistic neutron in-flight beta decay, is capable of simultaneously re-illuminating secondary electrons of tens of TeV and their inverse Compton photons at such distant and unconnected distances. 
\subsection{PeV neutron by photo-nuclear disruption}
Such a resurgence  and alignment of tens of PeV neutrons do correspond to the observed puzzling TeV gamma-ray signatures in SS433.   A quite similar process, capable of  ejecting PeV neutrons, could   be caused,  even at lower energy, by the photodisintegration of light and medium nuclei.  Such nuclear channels are ruled  by  the Giant Dipole Resonance, GDR.   These  reactions  are  a key processes in Ultra High Energy Cosmic Ray,  UHECR,   for their confinement.  The tens EeV ($10^{18}$ eV)  UHECR were discovered  to be mainly  made by lightest nuclei as He, Li, Be. Their GDR interaction with the cosmic $2.7$~K radiation make them very fragile and confined.  Within a few Mpc,  as the known  Local Sheet Galaxies in Local Group, \cite{fargion2024uhecr}.
For example, when a photon hits a light nucleus, such as Be (beryllium), with an energy higher than the separation energy, neutron emission prevails and a lighter Be isotope is produced: for Be the separation energy is 1.7 MeV, while the GDR is about ten MeV.
However, the fraction of energy in the final neutron is only a small part of the initial energy of the nucleus.  Therefore, it is less effective in forming a neutron beam than proton photopion conversion by the Delta resonance and its decay. These complex reaction chains may occur at lower photon energies than the main photo-pion reaction discussed below, which is ruled by the formation energy of the Delta resonance.  Such GDR events with nuclei are neither unique nor simple, nor useful for understanding the model. These possibilities might be considered in future works.

\subsection{Tens TeV gamma splitting onto  far W50   shell  nebula. An alternative model ?}
 In a different, very fine tuned  model, we may consider the possible presence in the same W50 nebula around SS433 of a  very peculiar   density in the W50 nebula  shell gas.  Tens of thousands of TeV gamma beams burst, which originated a century ago directly from SS433 may hit this thin shell layer,  split into electron pairs,  and re-emit again  TeV photons by inverse Comption scattering (ICS) \cite{fargion1998inverse}. Such a  peculiar target may act as an optimal beam dump scattering for several tens or hundreds of TeV photons ejected in a past SS433 highest jet event.
 Future observations by HAWC,  H.E.S.S. and LHAASO could be able to disentangle, validate or reject each of the two models. However, the presence of such bright TeV radiation all along tens or hundred light-year distances from SS433 makes this tuned nebula mass density  (not too dense to not be absorbed, nor too diluted to be not interacting) an ad-hoc model somehow prone to criticism.

\section{Microquasar, precessing Jets,  GRBs and recent  UHECR Amaterasu event}
Micro-quasars are binary systems where a NS of a few solar masses, or an heavier BH of several or tens solar masses, are bound by gravity with an orbiting star of a  comparable or larger mass, while capturing the star mass by tidal forces.  As the star's mass tail collapses onto the SN (or BH), the same mass fuels an accretion disk around the SN (or BH).  This rotating disk typically induces asymmetric charge flows and consequently enormous currents that create a powerful toroidal magnetic field. 

The magnetic field shrinkage  can undergo sudden and repetitive  temporal variations. These events also induce extreme pulsed spiral electric fields that accelerate charges along the disk's edge both hadrons and electron pairs. 
The electron at the outer edge of such disks can shine in twin rings  separated by the BH or NS,  as, for example, in the twin disk  presence observed in the Crab Nebula edges.
The inner free charges (positive  protons and nucleons, as well as negative electrons) spiral above and below the same accretion disk.  
These ultra-high-energy charges are quickly constrained and  collimated by the extreme magnetic fields: their narrow Larmor radius forces them to form a very  thin cylindrical cone ejected  along the twin poles.

\begin{figure}[ht]
  \centering
    \includegraphics[width=0.7\textwidth]{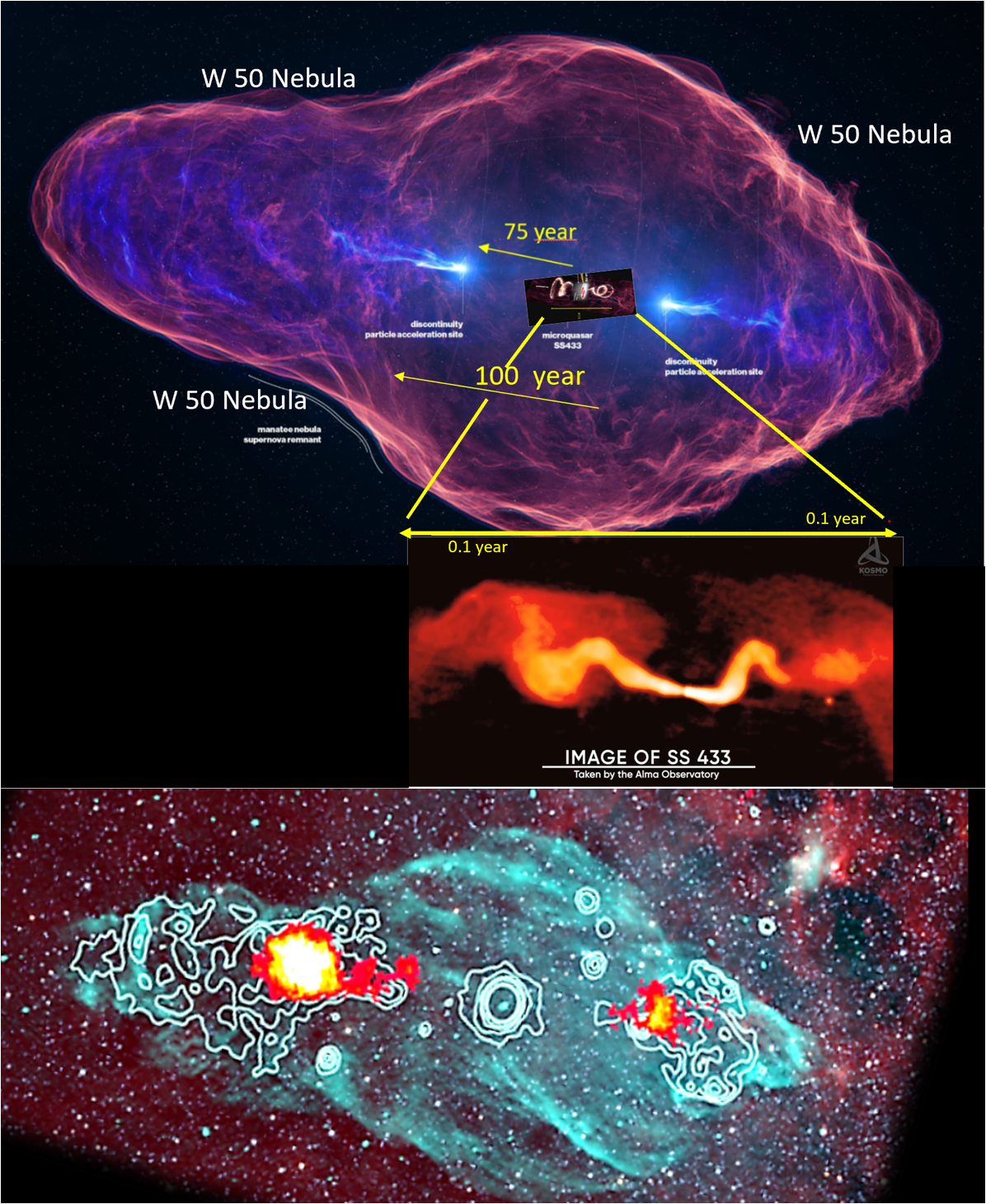} 
    \caption{
      Above an artist's impression of the SS 433 system, showing the jets (blue) and the surrounding W50 (red): the jets travel undetected for a distance of about 75 ly before suddenly reappearing as bright sources of non-thermal emission (X-rays and gamma rays) as observed by H.E.S.S., HAWC and LHAASO inside the nebula long before generated by a supernova.
      A view of the SS433 spiral jet at a millimetre wavelength is shown in the inset, taken by the radio telescope array ALMA.
      Below~\cite{hess2024acceleration} the image of the H.E.S.S. data with the additional contour of the radio radiation.
      The image of the collimated two-TeV gamma-ray beams, separated at a distance between 75 and 150 ly, is very puzzling.  It inspired  present PeV neutron  beam model.
     }
    \label{fig:1}
\end{figure}

Eventually, these relativistic particles, both hadrons and their attracted and  collimated electrons, are  ejected vertically in twin cones, aligned along the North or South axes of the NS or BH,  magnetic poles. 
Tidal forces among the orbital star companion and the accretion disk may drive the  SS433  jet into a  conical precession  jet. These tidal events, in their initial and final mass accretion phases, can feed a thin, persistent, precessing jet, whose intensity can decay over time after the destruction or collapse of the brighter  star companion.  Usually we observe the system off axis, as in SS433 case. see Fig \ref{fig:1}.
This thin,  spinning  jet can rarely be observed on the axis  pointing to us, like the distant and far cosmic gamma-ray burst (GRB)  or the much closer (more off axis)  and longer-lived soft gamma-ray repeaters (SGR)~\cite{fargion1999nature, fargion2006grbs}.

 The main processes that produce the observed gamma rays in GRBs should be the ICS  \cite{fargion1998inverse}. The leptonic component, mostly electron pairs in the jet, is also leading by synchrotron radiation to some radio, X-ray, rare optical, and gamma photons. The hardest gamma signals (MeV-GeV-TeV)  by ICS have been discovered since half a century and also, very recently, in surprising details along SS433. The GRB apparent huge luminosity is connected with the thin jet beaming in axis  to us  associated with their huge (thousands-millions) relativistic Lorentz factor; their variability is due to the jet spinning and precessing beam. The very recent and rarest  long life GRBs and their unexpected (and unexplained)  repetition \cite{levan2025daylongrepeatinggrb250702bde} is just an additional    argument for  a persistent and steady accelerating jet model as for GRBs  (as well as  for SGRs) \cite{fargion2006grbs}. 
 
 The SS433 system is somehow such a near, off axis,  example here to  be considered in detail, well before the same binary system  will finally collapse into a future more dramatic tidal collapse jet observable,  if in axis, as a GRB event. The tidal collapse of NS-NS,  NS-BH  or BH-BH  (with  live accretion disks) may also lead to  very rare  (because of the very narrow beam jet  alignment)  associated GRB-GW signal \cite{Fargion:2017uwg}.     
 Finally, the huge relativistic Lorentz factor acceleration needed in the tens of hundreds of PeV   range  for  a proton,   $10^7- 10^8$ GeV,  might be associated with a more extreme relativistic event,  possibly occurring at the explosive   W50  birth event,   nearly $20.000$ years ago forming the heavy Black Hole object.   In  that case, an earliest brightest supernova and GRB event was  possibly associated with the  later survived SS433  BH  system. It  could also be  ejecting   a maximal  jet  and a contemporaneous  UHECR event,   with a corresponding  relativistic Lorentz factor of  $10^9- 10^{10}$. This value is for heavy nuclei; for UHECR events, we have: $ E_\text{UHECR} > 10 ^{19} \text{eV}$ (for protons or lightest nuclei)  and  $ E_\text{UHECR} > 10 ^{20} \text{eV}$  for heavier nuclei. %
Consequently,  SS433, the closest galactic micro-quasar source, could be the  observed  source of a rare clustering  of four UHECR events,  whose presence around SS433 was  observed in the last decades by AUGER and TA~\cite{fargion2025lightest,fargion2024uhecr}.
Moreover, the possible heaviest UHECR nuclei, $Fe, Ni$, with a corresponding huge relativistic Lorentz factor of $10^8-10^{10}$, ejected in the W50 explosive event about 18,000 years ago, could  reach us with some delay (due to the bending curved trajectory)  and in a slightly deviated direction, due to the huge charges of the heavy nuclei and the  consequent  path  arc,  or random walk, \cite{fargion2011coherent} along  the galactic magnetic field.
This possibility  may explain the most energetic (and not yet understood) UHECR event named "Amaterasu", that is displaced by a few tens of degrees from SS433 direction.  An event observed four  years ago by TA~\cite{telescope2023extremely,fargion2024uhecr}  in a sky volume without any observed AGN candidate source. %

\section{  GZK cut-off, Delta resonance  and  SS433 separated TeV twin jet }
As the name suggests, micro-quasars are just a small scale (within a parsec fraction) system of the more famous and larger quasars. Quasars are powered by millions or billion solar masses of BH, hidden in galactic centers, also known as Active Galactic Nuclei (AGN).  These ones are able to eject much wider, longer, and harder hadron  (and associated lepton) jets, even in megaparsec sizes. As we noted, SS433 is a small rare binary system containing a supergiant star that is overflowing its Roche lobe with matter onto a nearby BH, within distances of hundreds of light-seconds. Both objects have a mass as large as ten solar ones. 
Its size is small, but its proximity to us makes it an ideal test for inspecting the jet formation. 

The LHAASO experiment, the largest km-square gamma array, recently studied the system~\cite{sudoh2020multiwavelength}. The TeV beam appearance far from the source was discovered recently by H.E.S.S. and the Cherenkov Telescope Array. Also HAWC, another large array in Mexico, revealed the disconnected TeV beams.  The presence and collimation in such far separated beams at far distances is
quite puzzling.  Most models require surprising high  shock wave re-acceleration
and unexplained re-collimation.  We shall not discuss them in this article.
Here we suggest, as mentioned,  a different possibility, that one century ago a rare  explosive flare episode in this micro-quasar may shine, at the same time, ultraviolet hot photons and ultra-relativistic energy (UHE) proton (or nuclei) in the jet.

These proton  photo-nuclear scattering in a hot thermal bath,  their interactions, could form  Delta $\Delta^{+}$
resonances. The consequent  $\Delta^{+}$ decay may feed PeV protons  and,  by decay, neutral pions as well as relativistic neutrons  with their comparable  charged pions. Such a phenomenon has an analogy in cosmic volumes and in thermal infrared big bang radiation and takes place between UHECR at  highest $6\cdot 10^{19}-10^{20}$  eV energy  interacting with  cosmic black body photon at  $2.7K$. 

This phenomenon is well known under the initials of the author \cite{greisen1966end, 1966Jeptl} as the possible ”GZK cut-off”  responsible for the maximal values of the UHECR  energy spectra. 
(Today, most of us believe that the suppression of UHECR is ruled  mainly by fragmentation of the lightest nuclei by GDR.) 
We suggest here similar processes happening in a  much smaller volume around SS433,   but within a more dense hot luminous photon bath,   by  a PeV  energetic proton-UV photon scattering. 
Indeed, the same Delta resonance may decay both in a relic proton and in  neutron secondaries. The tens PeV  neutron  may escape un-deflected as a collimated energetic neutron jet. The presence of such tens PeV neutron beams,  its non-radiating flight  path, and  its much later  separated decay  explain both TeV  beta and photon trace signals, as the cornerstone of our model. 
The observed 75 (up to hundreds) ly distance of the UHE neutron decay, the electron radiation resulting from the ICS as a tens of TeV gamma-ray reemerging signal, is the mechanism that solves, in our opinion,  the enigmatic and sudden reappearing TeV tracks of SS433.%

\section{Delta resonance \texorpdfstring{$\Delta ^+$}{D'} decaying  into neutrons}

 Recall that a neutron must exponentially decay within nearly 877 seconds.  Assuming the corresponding flight distance at a relativistic regime, one obtains:
\begin{eqnarray}
  L_n &=& 877 \; \left(E_n/m_n \right)    \;  \text{ls}\\
  L_n &=& 75  \; \left(E_n/ 25 PeV\right) \;  \text{ly}
\end{eqnarray}

 These 75 ly distances are the earliest staring disconnected TeV  track signal  from the SS433; its extension may be twice larger assuming twice the UHE neutron energy.  We first use it to fix a meter size for the model.
Let us recall the necessary tuned resonant condition  for $\Delta$ mass creation and photo-nuclear interaction energy
\begin{equation}
  m_{\Delta} = \sqrt{2\cdot E_p \cdot E_{\gamma}}
\end{equation}

The known $\Delta$ mass  is $m_{\Delta} = 1232\pm 2 $ MeV, and  the  expected $25 $  PeV proton  or neutron  energy define, as a first approximation,   a  corresponding tuned photon temperature:

\begin{equation}
 E_{\gamma} = \frac{\left(m _{\Delta}\right)^2 }{2 \cdot E_p} = 
  \frac{30.3\; \text{eV}}{\left(E_p/(25 \; \text{PeV})\right)}
\end{equation}
However, the final neutron or proton secondary of the ultra-relativistic decay $\Delta^+$ should also take into account the lost energy in the decay of the delta for  the companion charged-pion secondary.  Therefore, to reach  a final  $25$ PeV  neutron energy one must assume a  $10\%$ additional primary proton energy. This is  leading to a tuned necessary energy photon as follows: 
 \begin{equation}
  E_{\gamma} = \frac{\left(m _{\Delta}\right)^2 }{2\cdot E_p} = 
   \frac{27.6\; \text{eV}}{\left(E_p/(27.5 \; \text{PeV})\right)} =
   \frac{3.2\; 10^5 \text{K} \cdot k_{\mathrm{B}}}{\left(E_p/(27.5 \; \text{PeV})\right)}
\end{equation}

This thermal energy is in the ultraviolet range :  it is nearly  $55.4$  times  more energetic than our solar spectra. Now we need a geometrical frame to estimate the photon volume and number density  where the expected SS433  proton jet could interact, producing the delta baryon resonance $\Delta^+$. This condition is not, a-priori,  a guaranteed possible one.  

\begin{figure}[ht]
  \centering
    \includegraphics[width=0.5\textwidth]{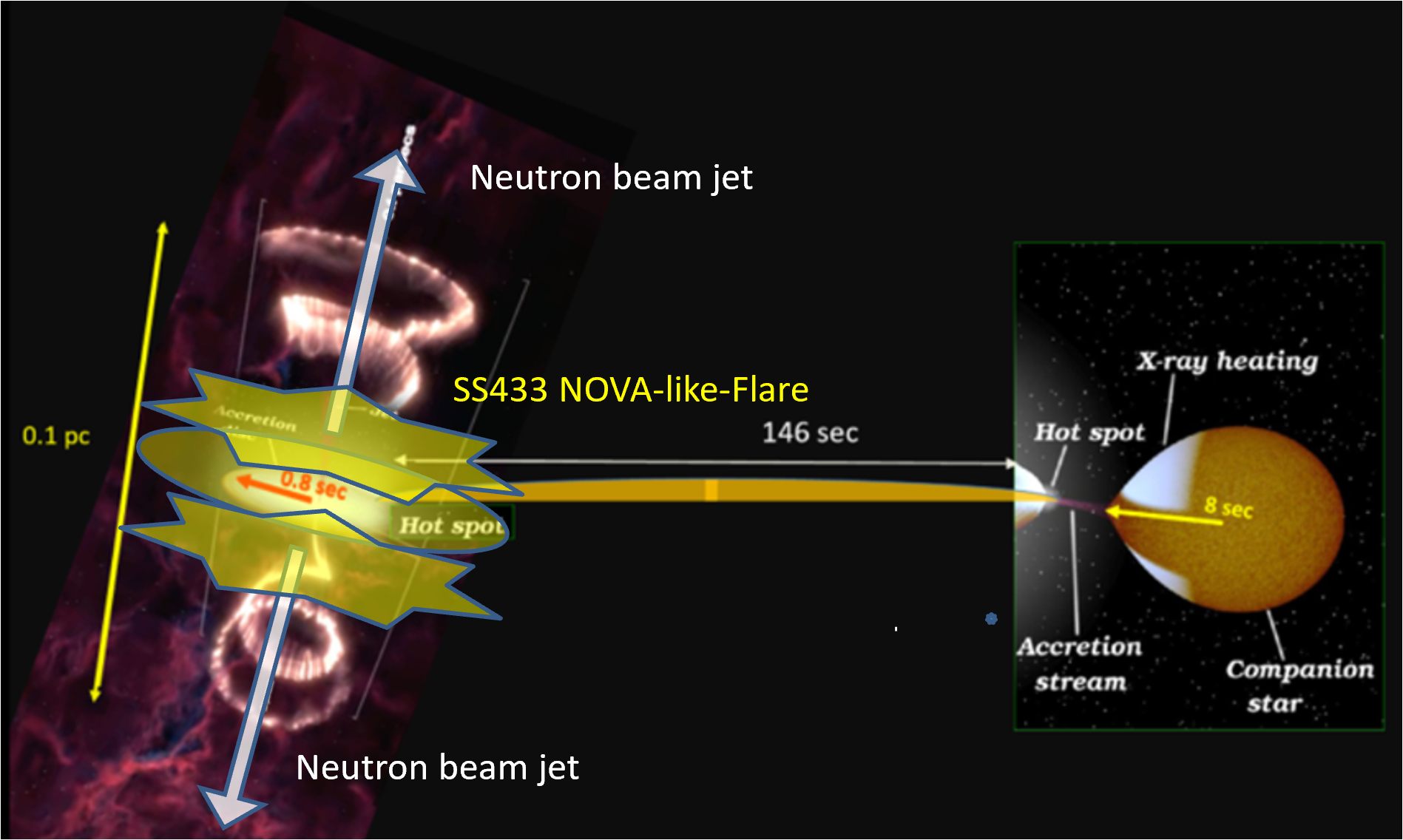} 
    \caption{A rough scale schematic of the inner binary system SS433: an accretion disk on a celestial body and a companion star of comparable mass, both nominally ten solar masses, at their respective Kepler circular distances. The radii of the star and, in particular, of the accretion disk are approximate for a complete view only, but could be slightly or significantly larger. We show here  the explosive stage, occurred possibly a century ago, when a Nova-like event offered the necessary luminosity to allow  tens PeV proton coversion into tens PeV neutron beam.
    }

    \label{fig:2}
\end{figure}

 \section{The   photo-nuclear conversion of proton into neutron}
 Let us estimate the present SS433 binary system size, as the BH companion distances, the most  probable accretion disk (whose brightening area  is the main photon volume  for the photo-nuclear event).  This estimate  may be offered, in the first approximation, by the known Kepler  orbit and by the most probable BH, and by a comparable  companion. 
 Their distance,  assuming a main  $M_\text{BH} = M_\text{Star}$ and $M_\text{BH}$ =10 $M_{\odot}$, and considering their observed Kepler periodicity, of the 13 day period (and the 162 day precession time) is:  

\begin{equation}
   D= 146 \cdot \left( \frac{M_\text{BH}}{10 \; M_{\odot}} + 
                       \frac{M_\text{Star}}{10 \; M_\odot} \right) ^{1/3} \; \text{ls}
\end{equation}

Note that such a result is quite stable, because the cubic root is  not much depending on the exact system masses.   Now we may estimate as a first evaluation the photon density in such a hot flare and temperature occurring at the accretion disk area. We here assume,  as a first  estimate,  an accretion disk area comparable with our solar one.  Therefore, the  Boltzmann law implies a  corresponding  flare luminosity  (respect  to our solar black body one) as large as the fourth power of the temperature ratio, or ${(55.4)}^{4} = 9.4 \cdot 10^{6}$,  corresponding to:

\begin{equation}
  L_\text{Flare}= 3.57 \cdot 10^ {40} \cdot \text{erg/s}
\end{equation}
This event energy (or higher ones) is comparable to that of a common nova flare.  The photon  number density in the same SS433  flare event is  ${(55.4)}^{3} = 1.7 \cdot 10^{5}$ the solar one  ($ n_{Th} = 3.9 \cdot 10^{12} cm^{-3}$). From the luminosity and the number density $n_{Th}$ as well as  from the following  Delta $\Delta^+$ resonant  peak  photo-nuclear resonance  cross section 

\begin{equation}
   \sigma_{\Delta} = 500 \;  \mu\text{b}  
\end{equation}
  We may derive  the probability  conversion 
\begin{equation}
    P[\gamma p \rightarrow  \Delta \rightarrow n+ \pi ] \propto   
    1-e^{- \sigma_{\Delta} n_{T} D_j}
\end{equation}

\begin{equation}
  \sigma_{\Delta}\cdot n_{Th} \cdot D_j = 23.2>>1 
\end{equation}

Here $D_j$  is assumed to be comparable with our solar radius.
Therefore, the necessary condition for an efficient proto-neutron conversion is quite well  satisfied.
It should be questioned whether the longer distant neutron beam  as far as  $150$~ly is also respected. 
The needed energy will be twice larger, the  tuned  flare energy and temperature  will be twice smaller, 
and the consequent $n_{Th}$ will be $8$ times smaller, as the same photo-pion probability %
estimate that decreases  as follows: 
\begin{equation*}
    \sigma_{\Delta}\cdot n_{Th} \cdot D_j = 2.9 >1.
\end{equation*}
Therefore,  the possibility of creating with a flare in SS433 able to convert a proton jet into a 27 or 54 PeV neutron jet beam is within the model possibility.

\subsection{Constrains and collimation of the jet by Larmor radius at PeV and TeV energy} 
There is the simple question about the proton secondaries of the Delta event considered above. 
They are charged and bent by galactic fields according to the following Larmor radius formula for the proton, $R_p$:

\begin{equation}
 R_p = (E_p/25 PeV)\cdot(B_g/(3\cdot \mu\text{G}))^{-1} \cdot 26.4 \cdot \text{ly}
\end{equation}
Therefore, they cannot rule the SS433 beam as long as hundreds of years in a collimated way. Moreover, the comparable tens of TeV spiraling electrons,  whose re-emission  shine at TeV energies,  are also very much constrained by their Larmor radius $R_e $:

 \begin{equation}
 R_e = (E_e/25 TeV)\cdot(B_g/(3\cdot \mu\text{G}))^{-1} \cdot 0.026.4 \cdot \text{ly}
\end{equation}
These very narrow distances imply that these secondaries cannot escape far from their primary hadrons; in our model, the main PeV neutron trace.

\section{Final conclusions}%
The PeV neutron decay in flight is a model that may explain the separated TeV gamma  trace in SS433 map.
The proposal has some analogy with a much earlier proposal: the  relativistic PeV-EeV tau decay in flight created by cosmic neutrino escaping mountains or Earth \cite{fargion2002discovering, fargion2004tau}.  It also reminds us of a more recent proposal for a recent HAWC trace of TeVs gamma signals, possibly due to cosmic ray  skimming and showering in muon and in their beta decay,  reaching us  from  the Sun  shadows \cite{fargion2025over}.   Similar signals may one day reach us as ultra-relativistic  muons (made by PeV muon neutrinos)  in exit from the lunar soil. Their beta decay in flight can also send us tens TeV electrons from the  Moon \cite{fargion2018signals}), leading  to similar tens TeV  gamma-like  air-showers inside or near the Moon shadows.   An exciting  future for neutrino astronomy.  
Therefore we  just imagined the equivalent  PeV neutron role and its beta decay  for the puzzling TeV separated beam in SS433.
A  large past flare in SS433 could be, by  its thermal bath and  by tens PeV proton jet,  the source of a 25 PeV neutron jet. This may explain the puzzling separated twin TeV gamma beam at a distance of 75~ly. The hidden PeV neutron flight has been revealed   only when   its beam is resurged  by its tens of TeV electron rise. TeV gamma signals were formed only at  a  far distance. These electrons   may be shining by synchrotron X rays and ICS radiation at TeV energies. Its collimation is evident with respect to the  other model as the standard model of shock wave processes. 

An early trigger was an explosive event, nearly 75--150 years ago (see Fig.\ref{fig:2}) when a powerful nova-like flare or burst, for a duration of few hours or a few days, occurred in the SS433 system. It may have shined almost in a fixed direction, not along any wide conical preceding jet volume as the inner one. This event happened nearly a century ago, that is around the first or second world War epoch.
Today, novae are observed at a rate of ten a year in our galaxy. Such a nova event could have escaped detection in early or late postwar times. Indeed, the same nature of SS433 was discovered much later, in 1977. It could be possible and worth it anyway to inspect the oldest astronomical photo-plate array in that direction and at those epochs, looking for such a sudden variability signal in luminosity.  The presence of such  PeV neutron-separated beam in SS433 may also suggest to search in other micro-quasar systems for similar disconnected signatures. 
Their statistics could define a corresponding rate of neutrino spectra in the energy range PeV to hundreds of TeV, with consequences for the ICECUBE records and neutrino flavors.
The 25 PeV neutron beta decay and its primary 27 PeV proton-pion event, while on axis toward us, may shine both of a brightest prompt ($1-2$) PeV gamma burst but also of a secondary (electron and muon) neutrino at ($0.2-0.5$) PeV.  These tuned energies are quite interesting: they define a critical energy for such neutrino events.  The energy window could be connected to the apparent TeV-PeV energy discontinuity in the ICECUBE neutrino spectra.

\section*{Acknowledgements}
The research of D.S. was performed in Southern Federal University with financial support of grant of Russian Science Foundation № 25-07-IF.
\bibliographystyle{JHEP}
\bibliography{daf2025v26}
\end{document}